\documentclass[useAMS]{mn2e}

\usepackage{graphicx}
\usepackage{bm}
\usepackage{psfrag}
\usepackage{mathrsfs}
\usepackage{amssymb,amsmath}

\newcommand{\be}{\begin{eqnarray} }
\newcommand{\ee}{ \end{eqnarray} }

\title[Radio properties of the Sgr A$^{*}$ magnetar]
{Radio properties of the magnetar near Sagittarius A$^{*}$
from observations with the Australia Telescope Compact Array.}

\author[R.~M.~Shannon \& S. Johnston]
{R.~M.~Shannon\thanks{Email: Ryan.Shannon@csiro.au} and S.~Johnston
\\
CSIRO Astronomy and Space Science, Australia Telescope National Facility, PO Box 76, Epping, NSW 1710, Australia}
\begin{document}
\maketitle
\begin{abstract}
We have carried out observations of the newly-discovered magnetar in 
the direction of Sagittarius A$^{*}$ using the Australia Telescope Compact Array
in four frequency bands from 4.5 to 20~GHz. Radio pulsations
are clearly detected at all frequencies. 
We measure the pulsar's dispersion measure to be $1650\pm50$~cm$^{-3}$pc,
the highest of any of the known pulsars.
Once Faraday rotation has been taken into account, the pulse profile is 
more than 80\% linearly polarized at all frequencies and has a small 
degree (5\%) of circular polarization.
The rotation measure of $-67000 \pm 500$~rad~m$^{-2}$ is the largest (in 
magnitude) ever measured for a pulsar but still a factor 8 smaller 
than Sgr A$^{*}$ itself.
The combination of the dispersion
and rotation measures implies an integrated magnetic field strength of
$-50$~$\mu$G along the line of sight.
The flux density appears to have increased by about a factor of two
between observations made 30 days apart.
This object therefore joins the small class of radio emitting magnetars.
\end{abstract}

\begin{keywords}
Pulsars: individual: PSR J1745--2900, ISM: Structure, Galaxy: centre  
\end{keywords}

\section{Introduction}
Magnetars are slowly-rotating neutron stars powered by the energy stored
in their strong magnetic fields rather than their rotational energy.
Typically magnetars have spin periods of $\sim$5~s, large spin down rates
of order $\sim$10$^{-11}$ and implied magnetic fields in excess
of 10$^{14}$~G. They have been discovered via their high energy emission
and their luminosities at X-ray energies and above exceed the spin-down
energy by an order of magnitude
(Duncan \& Thompson 1992, Beloborodov 2009).

To date, of the more than 20 magnetars now known, only
three have confirmed radio pulsations: 
XTE~1810--197 (Camilo et al. 2006), 1E1547--5408 (Camilo et al. 2008) and
PSR~J1622--4950 (Levin et al. 2010). The first
two sources were detected at radio wavelengths following high energy outbursts;
the latter source was first detected in the radio band.
All three radio magnetars share similar characteristics in that they show
complex pulse morphology, very high degree of polarization and a flat
spectral index, with PSR~J1622--4950 detected at frequencies
up to 24~GHz (Keith et al. 2011) and XTE~1810--197 detected as high
as 144~GHz (Camilo et al. 2007).

It has long been a goal of pulsar astronomers to find a pulsar in orbit
about the black hole at the Galactic Centre (e.g., Kramer et al. 2004).
In particular, even an isolated pulsar in a tight orbit around the
black hole has the potential to provide exquisite tests of general
relativity (Liu et al. 2012).
Although it is expected that pulsars should exist at the Galactic 
Centre (Pfahl \& Loeb 2004), the postulated high degree of scatter-broadening
(Lazio \& Cordes 1998) implies that searches need to be conducted
at high radio frequencies. In spite of many attempts however
(Johnston et al. 2006, Deneva et al. 2009, Macquart et al. 2010)
no radio pulsars have yet been found within
10 minutes of arc of the Galactic Centre.

The recent discovery of a magnetar in the direction of Sagittarius A$^{*}$
by the {\em Swift} satellite (Kennea et al. 2013) and the subsequent 
detection of pulsations by the {\em NuStar} satellite (Mori et al. 2013)
was therefore of great interest to the radio pulsar community and has been 
the subject of follow-up observations in the radio band with single dishes
(Burgay et al. 2013; Eatough et al. 2013a,b,c; Buttu et al. 2013). 
Here we report on interferometric observations made 
with the Australia Telescope Compact Array and discuss their implications.

\section{Observations}
Observations towards the Galactic Centre were made on two
separate occasions with the Australia Telescope Compact Array (ATCA), an
east-west synthesis telescope 
located near Narrabri, NSW, which consists of six 22-m antennas 
on a 6-km track. The configuration of the array was 6C
(shortest baseline 153~m, longest baseline 6000~m), maximising
the potential for point-source imaging. The observations took place
on 2013 May 1 starting at UT 12:30 and 2013 May 31 at UT 09:30.
The observing time was 10~hours and 4 hours respectively.

The ATCA has recently been equipped with broad-band feeds covering
the spectral region between 4 and 11 GHz. The new broad-band correlator
(CABB; described in Wilson et al. 2011) is capable of processing
4 frequency bands each with 2048~MHz bandwidth
with 1~MHz frequency resolution and full polarization capability.
In addition, it can perform `pulsar binning', allowing high time resolution 
construction of pulse profiles (Wilson et al. 2011).

The observations switched between 2 frequencies with a cadence
of 30 min.  We observed from 4.5 to 8.5~GHz and from 16.0 to 20.0~GHz
with 1024 spectral channels (i.e. 4~MHz spectral resolution) and 32 
phase bins (i.e. $\sim$117~ms time resolution) folded at the spin period
of $\sim$3.767~s given by the {\em NuStar} observations of Mori et al. (2013).

Data reduction and analysis were carried out with the
{\sc miriad} package\footnote{See http://www.atnf.csiro.au/computing/software/miriad} using standard techniques. 
After flagging bad data, the primary calibrator (PKS 1934--638) was used
for flux density and bandpass calibration and the secondary calibrators,
PKS~1741--312 and PKS~1714--336, were used to solve for antenna gains,
phases and polarization leakage terms in the low and high frequency
observations respectively.

\section{Results}

\subsection{Position}
The Galactic Centre contains highly complex regions of bright radio emission,
making it difficult to image given the sparse uv coverage of the
ATCA observations. However, the advantage of pulsar binning is that we can
create visibilities formed from the difference between the on-pulse and
off-pulse bins (using the {\sc miriad} task {\sc psrbl}).
This therefore removes all non time-varying emission 
from the visibilities, resulting in an image which contains only the 
point source pulsar.

We created such an image using multi-frequency synthesis of the data using
the May 1 observations between
16.0 and 18.0 GHz, the band and epoch in which the most precise 
measurement of the position of the pulsar could be made.
The resultant beam size is $0.46\times1.84$~arcsec.
The location of the pulsar, as determined with the {\sc miriad}
routine {\sc imfit} is (J2000) 17:45:40.164, --29:00:29.818
with measurement uncertainty of 0.022 and 0.090 arcsec, respectively. 
This position is consistent with, but has smaller error bars than the Chandra 
position reported by Rea et al. (2013).

\subsection{Pulse profile}
\begin{figure} 
\begin{tabular}{c}
\includegraphics[scale=0.5]{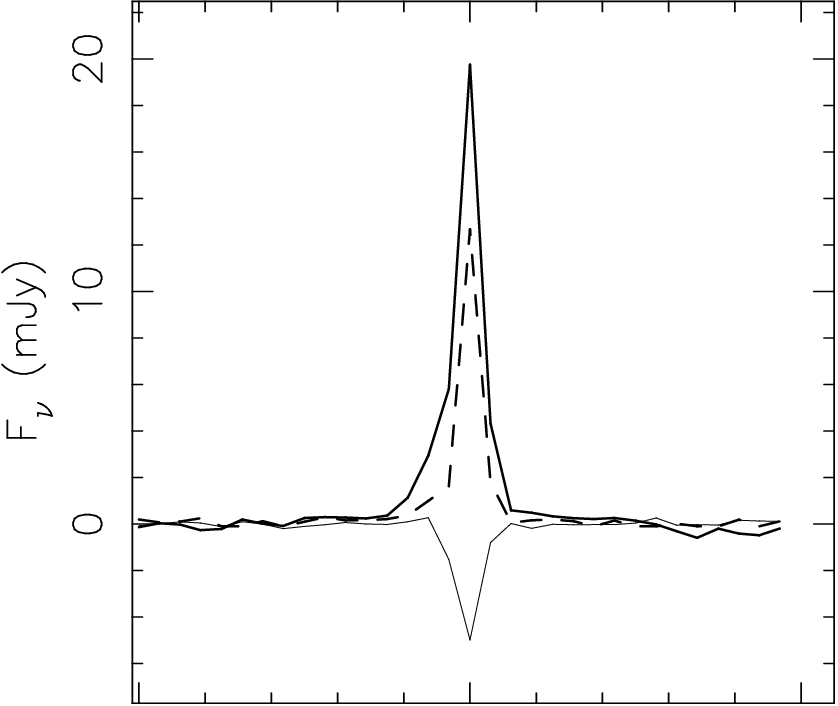}  \\
\includegraphics[scale=0.5]{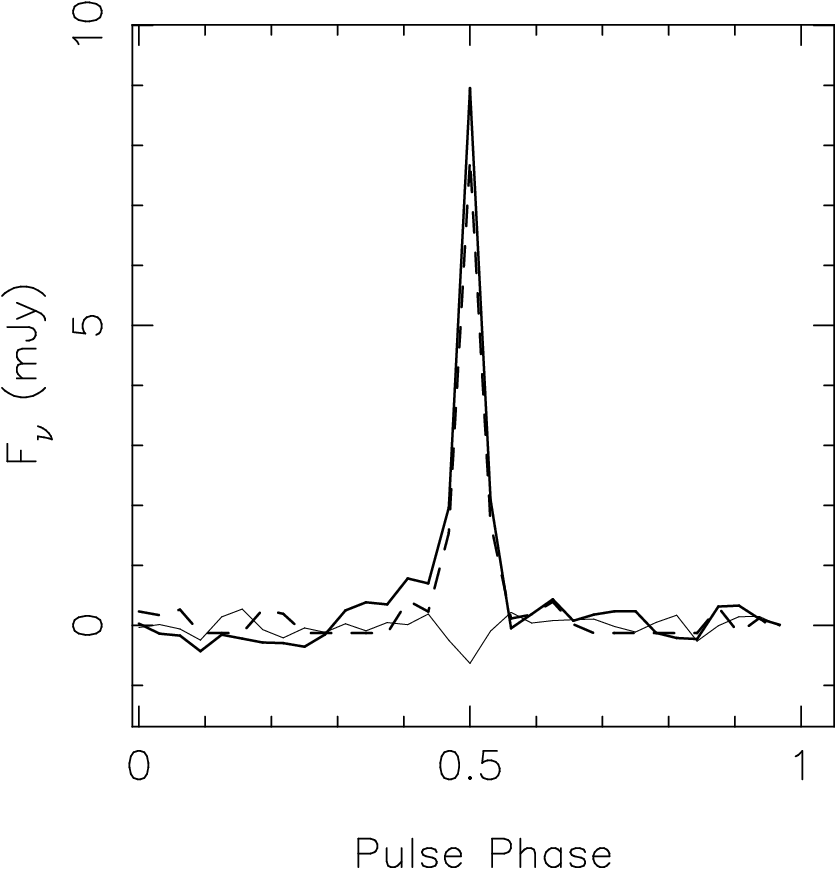}  \\
\end{tabular}
\caption{Pulse profile of the magnetar PSR~1745--2900 at 5~GHz
(top panel) and 17~GHz (lower panel). Solid line denotes total
intensity, dashed line is linear polarization and lighter solid
line is circular polarization.}
\label{fg:profile}
\end{figure}

At all frequencies, the pulsar is highly linearly polarized (see below)
and possesses some degree of circular polarization.
In the 4.5 to 8.5~GHz band, the pulse has a shallow rising 
edge and fast falling edge,
reminiscent of the trailing partial cones identified in 
Lyne \& Manchester (1998). The linear polarized fraction as 
measured is only 55\% and the circular polarization is 22\%.
However, even across the 4~MHz channels there is a significant
fraction of depolarization due to the high rotation measure.
At the higher frequency there is evidence also for a weak
trailing component to the pulse. The fractional circular polarization
is lower at 5\% but the linear polarized remains above 80\%.
It is difficult to draw any further conclusions given the
low time resolution of our data.
Pulse profiles are shown in Fig~\ref{fg:profile}.

\subsection{Dispersion Measure (DM)}
Although our temporal resolution is rather crude with $\sim$117~ms per
phase bin it is still possible to estimate the DM based on the 
comparison between the observations made over the frequency range
between 4.5 and 8.5~GHz.  Between these two frequencies the delay
difference amounts to $\sim$15~ms per 100 DM units.
We cross correlated pulse profiles made in 16 frequency bands between
4.5 and 8.5~GHz and determined the delay to be close to
2.0 phase bins. This implies a DM of 1650$\pm$50~cm$^{-3}$pc, consistent with
but more precise than the value reported by Eatough et al. (2013b).

\subsection{Polarization and Rotation Measure (RM)}
If the pulsar is in the Galactic centre region the electron density and 
magnetic field are expected to be high. 
We  therefore searched for rotation measure using the 17~GHz data because 
at this frequency the pulsar is bright and the effects of Faraday rotation 
and interstellar scattering are smaller.  We divided the 2~GHz bandwidth
between 16 and 18~GHz into 16~MHz sub-bands. 
\begin{figure} 
\includegraphics[scale=0.5]{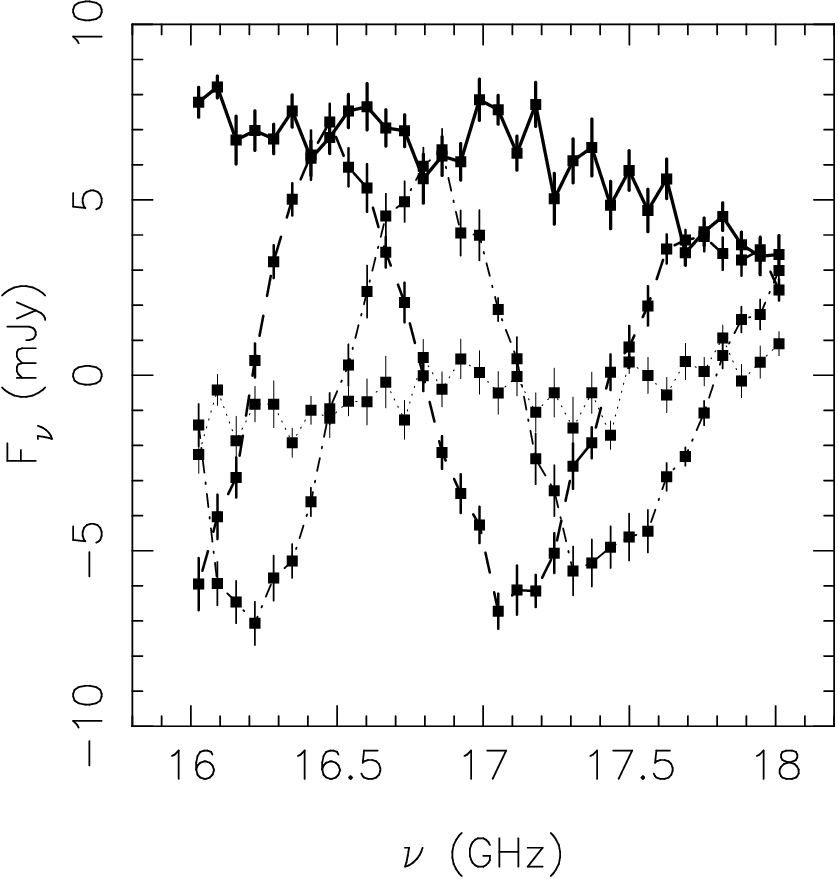}  \\
\caption{Flux density of the stokes parameters in the peak bin of
the profile as a function of observing frequency. The thick solid line 
shows Stokes I, Stokes Q and U are the 
dashed and dash-dotted lines, and Stokes V the thin solid line.}
\label{fg:iquv}
\end{figure}

\begin{figure}
\includegraphics[scale=0.5]{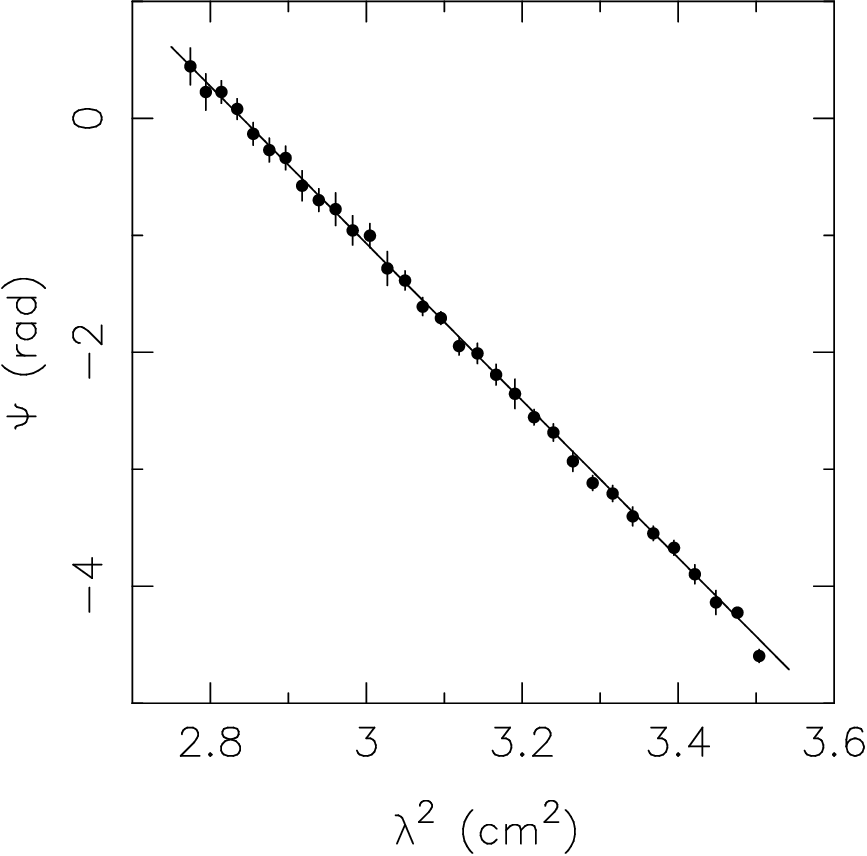}  \\
\caption{Position angle ($\Psi$) of the linear polarization plotted against 
the square of the observing wavelength  (at observing frequencies between 
16 and 18 GHz) for the peak bin in the profile.  The line shows the fit to the 
data; the best-fit RM is $-67000 \pm 500$~rad~m$^{-2}$.}
\label{fg:rm}
\end{figure}

Visual inspection of the pulse profiles showed a high degree of 
linear polarization in the individual frequency channels and clearly showed 
rotation of Stokes Q and U between channels.
Figure~\ref{fg:iquv} shows the Stokes parameters as a function of
observing frequency measured at the profile peak using the {\sc miriad}
function {\sc psrplt}. Two things can be noted from the figure. First,
the fractional polarization is very high, approaching 100\%.
Secondly, there is a large rotation of Stokes Q and U, implying a high
RM. We therefore computed the position angle ($\psi$) of the 
linear polarization as a function of observing frequency via
\begin{equation}
\psi  = 0.5 \,\,\, {\rm tan}^{-1} (U/Q).
\end{equation}
Figure~\ref{fg:rm} shows $\psi$ as a function of the square of the
observing wavelength. The RM is determined from a linear fit
to the data indicating an exceptionally high
value of $-67000 \pm 500$ rad~m$^{-2}$.

The extremely high rotation measure explains the non-detection of 
polarization reported by Eatough et al. (2013b) from Effelsberg 
observations, because it causes depolarization of signals if 
averaged over wide bandwidths.

\subsection{Flux density and spectral index}
\begin{figure} 
\includegraphics[scale=0.5]{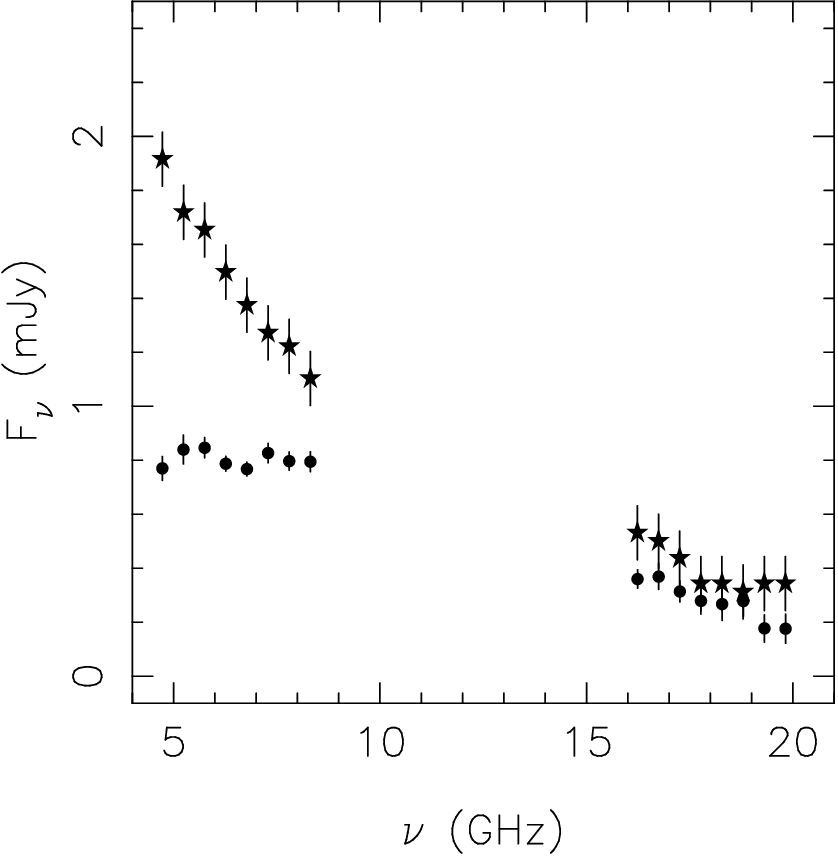}  \\
%\centerline{\psfig{file=PA.ps,width=\textwidth,angle=270}} 
\caption{Integrated (continuum) flux density $F_\nu$ of the magnetar as a 
function of the observing frequency $\nu$. Circles denote the May 1
observations, stars the May 31 observations.}
\label{fg:flux}
\end{figure}

We measured the flux density of the pulsar across four frequency bands
from 4.5 to 8.5~GHz and 16 to 20~GHz frequencies using the
{\sc miriad} task {\sc imfit} at each of the two observing epochs.
Figure~\ref{fg:flux} shows the results.

For the first set of observations,
the flux density at 6.5~GHz is 0.8~mJy and at 18~GHz is 0.3~mJy.
Between 4.5 and 8.5~GHz the flux densities are constant within the
error bars, yielding a spectral index of zero. The drop off in 
flux density between the 6.5 and 18~GHz would seem to imply a
spectral index of $-1.0$. However, we see an apparent steep flux
density decrease through the 16 to 20~GHz band.
Also, our measured values not easily reconciled with those
reported in Eatough et al. (2013c) which show a virtually flat 
spectral index between 2.4 and 19~GHz. Our values are significantly
higher at low frequencies and broadly consistent at the high
frequencies.
During the second set of observations the magnetar was considerably
brighter by about a factor two between 16 and 20~GHz. Futhermore,
unlike the earlier observations, the data show a steep decline 
in flux density between 4.5 and
8.5~GHz with a spectral index close to $-1.0$.

Magnetars are known to have variable flux densities
as a function of time (e.g., Levin et al. 2011), which may explain 
the differences in the two sets measurements and the differences
between our values and those in Eatough et al. (2013c).
We can rule out scintillation as a cause. The scintillation bandwidth
is extremely small, of order a few kHz at the low band and less than
100~kHz even in the higher band, well below our frequency resolution.
We also see no sign of variability {\it within} each observing epoch.

\section{Discussion}
The pulsar's angular proximity to the Galactic Centre coupled with
its high dispersion measure and large absorption in the X-ray band
make it likely that it is indeed very close to the centre of our
Galaxy. It is offset by 3 arcsec from Sgr A$^{*}$, implying a lower
limit to the physical separation of only 0.1~pc. 
The pulsar shares characteristics with the other known radio magnetars.
It has a high degree of linear polarization and remains bright at
high observing frequencies. It may also have variable flux density.

The measurement of both dispersion and rotation measures allows us
measure the average parallel magnetic field strength in
along the line of sight to the pulsar via $B_{||}$=1.2~RM/DM
with RM and DM in conventional units, and $B_{||}$ measured in $\mu$G.
This value is $-$50~$\mu$G, about an order of magnitude larger than 
typical line of sight measurements in the Galactic plane.
We note that the RM of the magnetar has the same sign and is about
a factor 8 smaller than the RM of Sgr A$^{*}$ itself (Bower et al. 2003)
but significantly larger than values of the non-thermal filaments.
Bower et al. (2003) argue that the RM does not originate from the
scattering screen but rather from material closer to Sgr A$^{*}$.
This would then imply that the magnetar is within this same material
but in front of Sgr A$^{*}$ as seen from Earth.

The nature of the magnetic field in the Galactic Centre is unclear.
Synchrotron filaments are present in the inner region with mG 
magnetic fields, but it remains the topic of debate as to whether fields
this strong are typical of the region as a whole. Indeed current
ideas indicate weaker fields of order 100~$\mu$G might be more typical
(Boldyrev \& Yusef-Zadeh 2006). Indeed, Crocker et al. (2010) claim that
the inner few hundred pc of the Galaxy has a field strength of order
50 - 100~$\mu$G, in line with our measurement. 
Continued monitoring of the source will will yield rotation measure and 
dispersion measure variations, enabling improved characterization of the 
interstellar medium and magnetic field structure in the Galactic Centre region. 

If the physical separation from the Galactic Centre is indeed 0.1~pc,
then the orbital period is greater than 1500~yr assuming
a circular orbit about Sgr A$^{*}$. This orbital period is too long to 
provide interesting constraints on gravitational theories (Liu et al. 2012).
We note that the pulsar would easily have been
detectable (if on at the time) in all the targetted surveys of the 
Galactic Centre carried out to date
(Johnston et al. 2006, Deneva et al. 2009, Macquart et al. 2010).
Although we cannot measure the scattering from our data, it appears
to be low - only a few ms at 8~GHz (Eatough et al. 2013c).
This gives optimism that observing frequencies around 10~GHz
might well be optimal for future surveys. Continuing searches of the 
Galactic Centre for pulsars are therefore warranted.

\section*{Acknowledgements}
The Australia Telescope Compact Array is part of the Australia Telescope,
which is funded by the Commonwealth of Australia for operation as 
a National Facility managed by CSIRO.

%\bibliographystyle{mn2e}
%\bibliography{journals,modrefs,psrrefs,crossrefs}
\end{document}